\documentclass[conference]{IEEEtran}
\IEEEoverridecommandlockouts
\usepackage{cite}
\usepackage{amsmath,amssymb,amsfonts}
\usepackage{algorithmic}
\usepackage{graphicx}
\usepackage{textcomp}
\usepackage{xcolor}
\def\BibTeX{{\rm B\kern-.05em{\sc i\kern-.025em b}\kern-.08em
    T\kern-.1667em\lower.7ex\hbox{E}\kern-.125emX}}
\begin{document}

\title{Self-Adaptive Large Language Model (LLM)-Based Multiagent Systems }
%to Support Human-Machine Teaming
% LLM-based adaptation 

\author{
	\IEEEauthorblockN{Nathalia Nascimento, Paulo Alencar, Donald Cowan}
	\IEEEauthorblockA{\textit{David R. Cheriton School of Computer Science} \\
		\textit{University of Waterloo (UW)}\\
		Waterloo, Canada \\
		\{nmoraesd, palencar, dcowan\} @uwaterloo.ca}
}

\maketitle

\begin{abstract}
In autonomic computing, self-adaptation has been proposed as a fundamental paradigm to manage the complexity of multiagent systems (MASs). This achieved by extending a system with support to monitor and adapt itself to achieve specific concerns of interest. 
Communication in these systems is key given that in scenarios involving agent interaction, it enhances cooperation and reduces coordination challenges by enabling direct, clear information exchange. 
However, improving the expressiveness of the interaction communication with MASs is not without challenges. In this sense, the interplay between self-adaptive  systems and effective communication is crucial for future MAS advancements.
In this paper, we propose the integration of 
 large language models (LLMs) such as GPT-based technologies into multiagent systems. 
We anchor our methodology on the MAPE-K model, which is renowned for its robust support in monitoring, analyzing, planning, and executing system adaptations in response to dynamic environments. We also present
a practical illustration of the proposed approach, in which we implement and assess a basic MAS-based application. The approach significantly advances the state-of-the-art of self-adaptive systems by proposing a new paradigm for MAS self-adaptation of autonomous systems based on LLM capabilities. 
%Progress in autonomic computing has lead to emergence of novel paradigms of collaboration such as human-machine teaming (HTM). In general, this paradigm relies on potential interactions, partnerships, and teamwork between humans and machines, thereby taking advantage of what each can do better. However, there are fundamental challenges in achieving HMT, including the need for increased transparency in human-machine interactions and augmented cognition to keep humans informed, support mutual exploration, and assist required adaptations. Further, it is not clear how multiagent systems can be combined with Large Language Models such as Chat-GPT to address these challenges.
\end{abstract}


\section{Introduction}

% Communication systems play an indispensable role in the success and efficiency of multiagent systems, enhancing the effectiveness of both software and embedded agents.  Regardless of the task's complexity, these systems harness communication structures that vary from rudimentary to sophisticated. For instance, even the most simple agents can employ elementary communication methods to coordinate their collective behavior for fundamental tasks like foraging, exploration, or assembly \cite{pagello1999cooperative}. On the other hand, more complex tasks, such as the development of autonomous robots or innovative communication strategies within agent populations \cite{marocco2007emergence}, often stimulate the emergence of adaptive communication systems \cite{sendra2023emergence}. This emergence stimulates self-organizing agent behaviors, predominantly evident through explicit communication strategies, where agents resort to clear, direct message passing to collaborate towards shared objectives. In advanced scenarios demanding human interaction, these robust communication systems prove invaluable. They enable agents to share feedback and decision-making information, thus fortifying human-machine cooperation and reducing potential coordination challenges \cite{cleland2022extending}.

In autonomic computing, the development of self-adaptive multiagent systems (MASs) is known to be a complex task \cite{fakhir2023smacs}. Self-adaptation is a well-known approach used to manage the complexity of these systems as it extends a system with support to monitor and adapt itself to achieve a concern of interest \cite{weyns2009self}. For example, by adjusting to changing scenarios, these systems can  optimize resource allocation or become fault tolerant by expressing high-level goals as utility functions. Communication is key in this regard. Even with basic communication constructs, simple agents can develop robust collective behaviors \cite{pagello1999cooperative} \cite{sendra2023emergence}. Conversely, complex tasks often trigger the emergence of adaptive behaviour, leading to self-organized, collaborative agents. In advanced scenarios involving agent interaction, these communication systems enhance cooperation and reduce coordination challenges by enabling direct, clear information exchange \cite{cleland2022extending}. The interplay of self-adaptive systems and effective communication is crucial for future autonomic MAS advancements.

However, improving the expressiveness of the interaction communication with MASs is not without challenges. 
%approaches . s  the design of MASs
%expressiveness of interactions communication. 
%However, the design of communication systems is not without its challenges. 
The increased complexity of these systems introduces synchronization overheads, thus necessitating careful selection of the approach best suited to the problem at hand \cite{altshuler2023recent}. This has led researchers to often opt for simple communication constructs, allowing robots to independently develop their own communication structures to address specific issues. Despite the inherent limitations of such an approach, the rapid advancement of Large Language Models (LLMs) and General Purpose Technologies (GPTs)\cite{zhao2023survey} \cite{wei2022emergent} \cite{huang2022large} provides a silver lining. These generative AI-based technologies allow for the integration of highly advanced conversational communication systems into software or hardware agents while using fewer resources.

In this paper, we propose a paradigm that integrates large language models (LLMs) such as GPT-based technologies into multiagent systems. By exploiting the rich capabilities of these advanced communication systems, we delve into the hypothesis of equipping autonomous agents with more sophisticated tools from the onset. We are particularly interested in the emergent abilities and capabilities these agents may exhibit when pre-equipped with such a powerful communication system. The primary input to these agents would consist of sensor data and communication from neighboring agents. In comparison with our prior approaches where agents evolved their own communication systems through evolutionary neural network algorithms \cite{do2017fiot}, the possibility we are exploring is of a paradigm shift in the agent's capabilities from the very inception. Will these agents still need to evolve and adapt their communication methods or will they be ready to execute complex tasks, leveraging the advanced communication systems inherent in the LLMs?

In our work, we present an innovative approach for developing self-adaptive agents using large language models (LLMs) within multi-agent systems (MASs). We anchor our methodology on the MAPE-K model, which is renowned for its robust support in monitoring, analyzing, planning, and executing system adaptations in response to dynamic environments. With this, we integrate GPT-4 technology, a cutting-edge LLM, enabling agents to adapt to more complex tasks and react to evolving situations intelligently. This, in turn, empowers our agents with improved communicative capabilities and adaptability.

The paper is structured as follows. Section 2 provides some research background and related work. Section 3 presents our approach, which relies on an LLM-based  Mape-K model. Section 4 presents a practical illustration of our approach, in which we implement and assess a basic MAS-based application. This experiment, presented in Section 3, exemplifies the application of our proposed approach. Section 5 concludes with summary remarks and future perspectives.

\section{Background and Related Work}

\subsection{LLM and GPT}
%LLM- o que ta por tras da estrutura, de forma q permita q ele seja o analyze, plan e o knowledge. 

Language Models (LLM) and Generative Pretrained Transformers (GPT) are integral parts of AI's Natural Language Processing (NLP) realm. While LLM is a broad category encompassing models that predict word sequences and can be used for various tasks such as text generation and translation, GPT, developed by OpenAI, is a specific LLM type. GPT, renowned for generating text akin to human writing, undergoes extensive pre-training before fine-tuning for specialized tasks. In essence, GPT is a subclass of LLMs, but not all LLMs are GPT models. Other prominent LLM examples include BERT, RoBERTa, and XLNet.

A GPT solution comprises several key components, such as a pretrained neural network model, a fine-tuning component to improve the model for specific tasks, an inference engine that uses the fine-tuned GPT model to generate responses or predictions (i.e. the inference engine feeds input data into the model and processes the model's output), and data pipeline that handles the flow of data in and out of the model \cite{brown2020language}. 

\subsection{Self-adaptive Systems: MAPE-K control loop}

The IBM control loop \cite{redbooks2004practical}, introduced in 2004, is a well-known architecture \cite{porter2020survey} for fostering autonomy and self-awareness in systems. The loop's framework, referred to as MAPE-K (Monitoring, Analyzing, Planning, Executing, and Knowledge), serves as a foundation for expanding self-adaptive and self-organizing systems \cite{do2017fiot}. The Monitoring stage involves collecting and associating data from the system's environment using specialized sensory functions. The Analyzing phase follows, where this monitored data is evaluated to determine necessary responses based on the environmental changes detected. Next, in the Planning stage, this analysis is used to narrow down a specific set of actions intended to reach a desired state within the system. Finally, the chosen actions are implemented in the Executing stage via effectors.

\begin{figure}[htb!]
	\centering
	\includegraphics[scale=0.40]{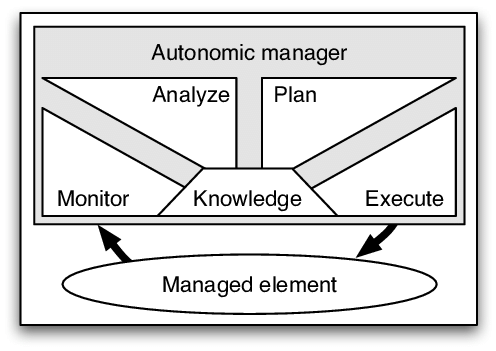} 
	\centering
	\caption{Original MAPE-K control loop to generate autonomous systems (Jacob et al., 2004), p. 24 \cite{redbooks2004practical}.}
	\label{fig:theory1}
\end{figure}

Several researchers have suggested incorporating the MAPE-K loop into multiagent systems \cite{farahani2017enabling}\cite{cleland2022extending} and have developed novel autonomic methods, either integrating or modifying the MAPE-K structure \cite{andersson2023architecting}\cite{dehraj2021review}\cite{do2017fiot}\cite{araujo2021lightweight}. Nascimento and Lucena, for instance, proposed substituting the 'analyze' and 'plan' stages with a neural network. In their model, sensor inputs feed the neural network, which in turn informs the agent's effector. The MAPE-K loop serves as a benchmark in this field.

% \subsection{Integrating LLM into MAS}
% %The integration of large language models (LLMs) into multiagent systems (MASs) is a relatively novel concept with a limited but promising body of related work. 
% Integrating Large Language Models (LLMs) like GPT-3 or GPT-4 into multiagent systems is a novel and emerging field. One of the few researchers that promote this integration is Feldt et al. \cite{feldt2023towards}, which explores using large language models (LLMs) as autonomous testing assistants in software development, offering a taxonomy and practical benefits.

\section{Approach: LLM-based MAPE-K Model}

% - Mostrar que a figura do MAPE-K do self-adaptive agent tem o LLM-based technology embutido nele, substituindo Analyze, Plan e Knowledge. 

In our research, we introduce an innovative architecture that integrates LLMs, specifically GPT-4, into multi-agent systems (MASs). Each agent within the MAS employs this technology in its control loop, creating an environment where every autonomous entity communicates and self-adapts using natural language processing. Our methodology is grounded in an extension of the MAPE-K model, renowned for facilitating adaptivity in dynamically changing environments. 

\begin{figure*}[ht!]
	\centering
	\includegraphics[scale=0.50]{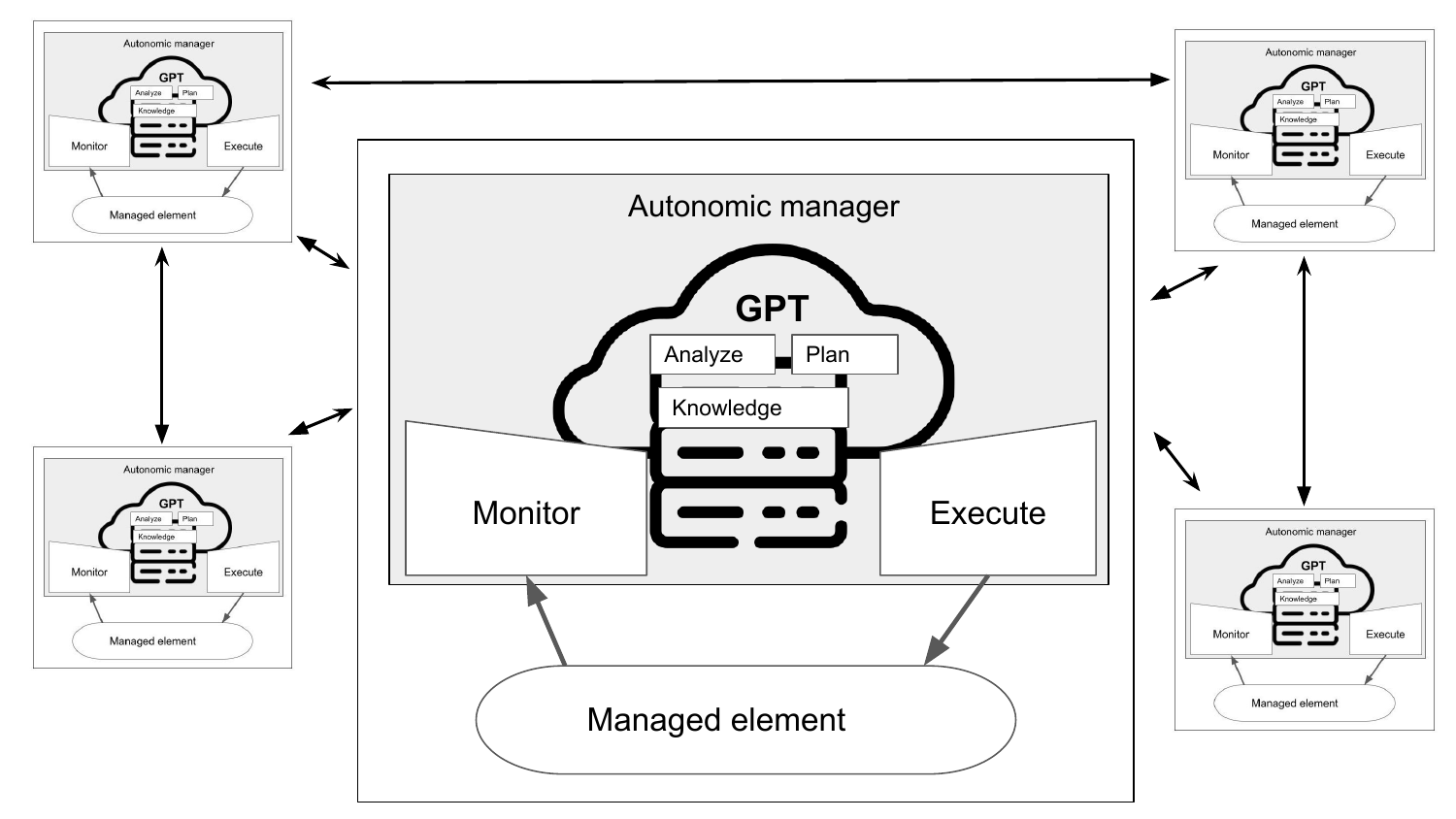} 
	\centering
	\caption{A multiagent system with self-adaptive autonomic agents. Each agent is self-managed through an adapted MAPE-K control loop.}
	\label{fig:approach}
\end{figure*}

As depicted in Figure \ref{fig:approach}, our proposed architecture modifies the traditional MAPE-K model, integrating the GPT-4, a state-of-the-art LLM, into the agent's control loop, enabling agents to adapt to and execute complex tasks while exhibiting advanced communication capabilities. This figure represents a MAS where each agent is autonomously managed through our adapted MAPE-K loop, comprising two core components: the managed element and the autonomic agent. 

The managed element comprises the environment with which the agent interacts, encompassing a range of sensors and actuators that monitor and control environmental elements. For instance, in a smart traffic application scenario, the managed element includes the monitored environmental factors (e.g., the number of cars and pedestrians) and the elements controllable by the agent (e.g., traffic lights). 

\begin{figure}[ht!]
	\centering
	\includegraphics[scale=0.47]{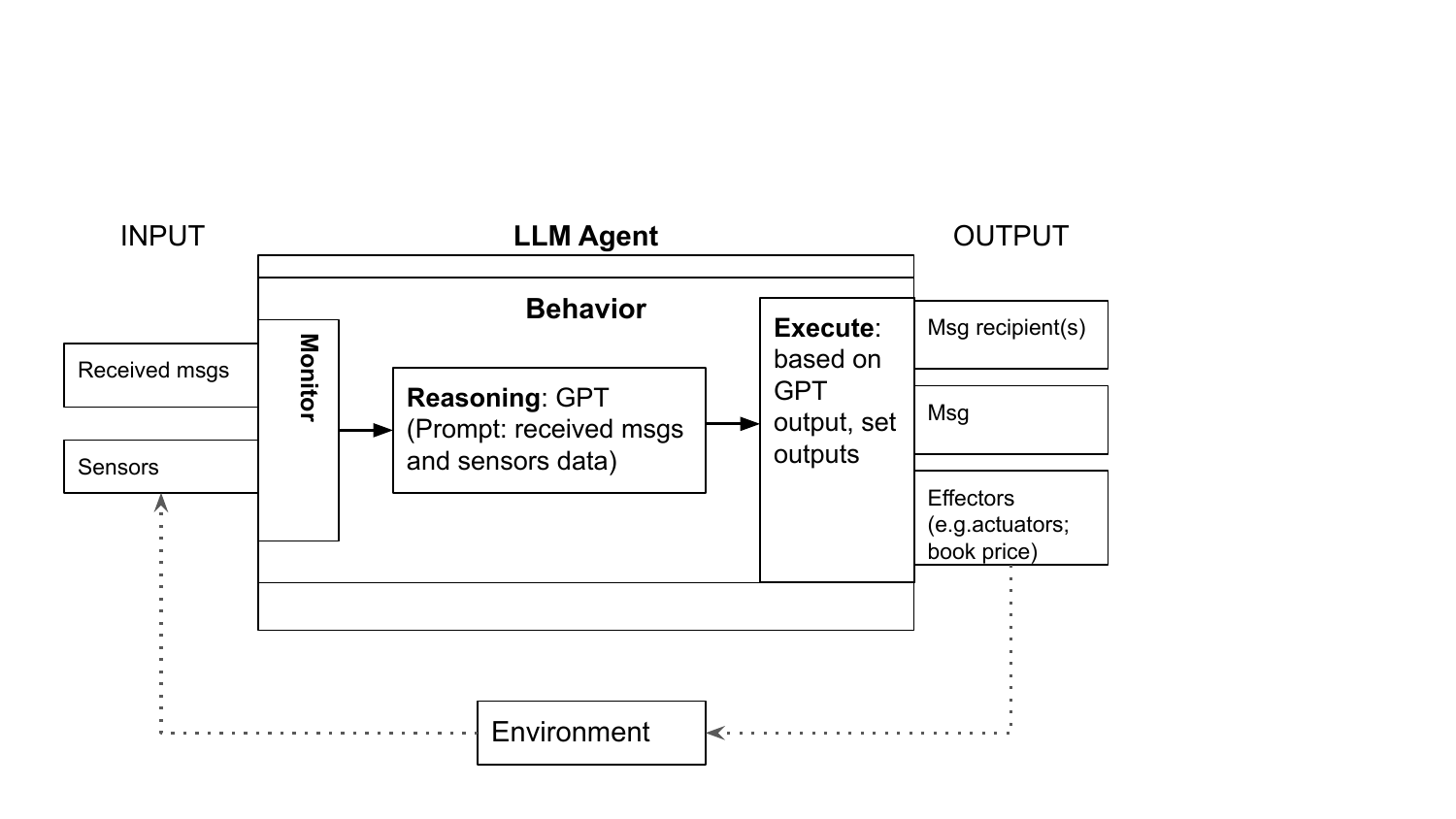} 
	\centering
	\caption{LLM-based agent.}
	\label{fig:llmagent}
\end{figure}

The autonomic agent, which is represented with more details in Figure \ref{fig:llmagent}, performs three primary tasks: 1) Monitor - this process collects data from the agent's sensors, processes the current state of the agent, and compiles messages from other agents. The consolidated information is transformed into a GPT-compatible prompt. If the agent receives messages from multiple agents, these messages are concatenated into a single prompt for each iteration; 2) GPT - this phase encapsulates the activities of analyze, plan, and knowledge. It operates the fine-tuned GPT model, with the pretrained neural network model and inference engine, to generate responses or predictions, handling the data flow in and out of the model; and 3) Execute - the GPT model's output is translated into an actionable command for the agent.

The intriguing aspect of this approach is its inherent adaptability. Each agent not only responds effectively to changes within its environment but also benefits from the advanced analytical capabilities of GPT-4. With LLM embedded into each agent, we posit that unique behaviors might emerge from such MAS. Therefore, our aim is to delve into the exploration of the potential behaviors  within these LLM-embedded autonomous agents as they interact and self-adapt.

% This architecture is designed for versatile adaptability, allowing each agent to effectively respond to evolving situations within their environment while leveraging the analytical and predictive power of GPT-4.

% - Cada agente tem um LLM embutido (o LLM faz parte do modelo do agent) 
% Integrating LLM to Autonomous Software Agents: // How powerful is their bargain capability? // Can They self-manage/self-adapt/self-organize through Natural Language? 
% mu

\section{APPLICATION SCENARIO}
\label{section:app}
In order to validate our approach for developing self-adaptive agents that leverage large language models (LLMs) within multiagent systems, we constructed a simple yet illustrative multiagent application. Our scenario, inspired by conventional examples found in multi-agent systems literature, consists of an online book marketplace, where autonomous agents act as buyers and sellers on behalf of users.

\begin{figure*}[htb!]
	\centering
	\includegraphics[scale=0.65]{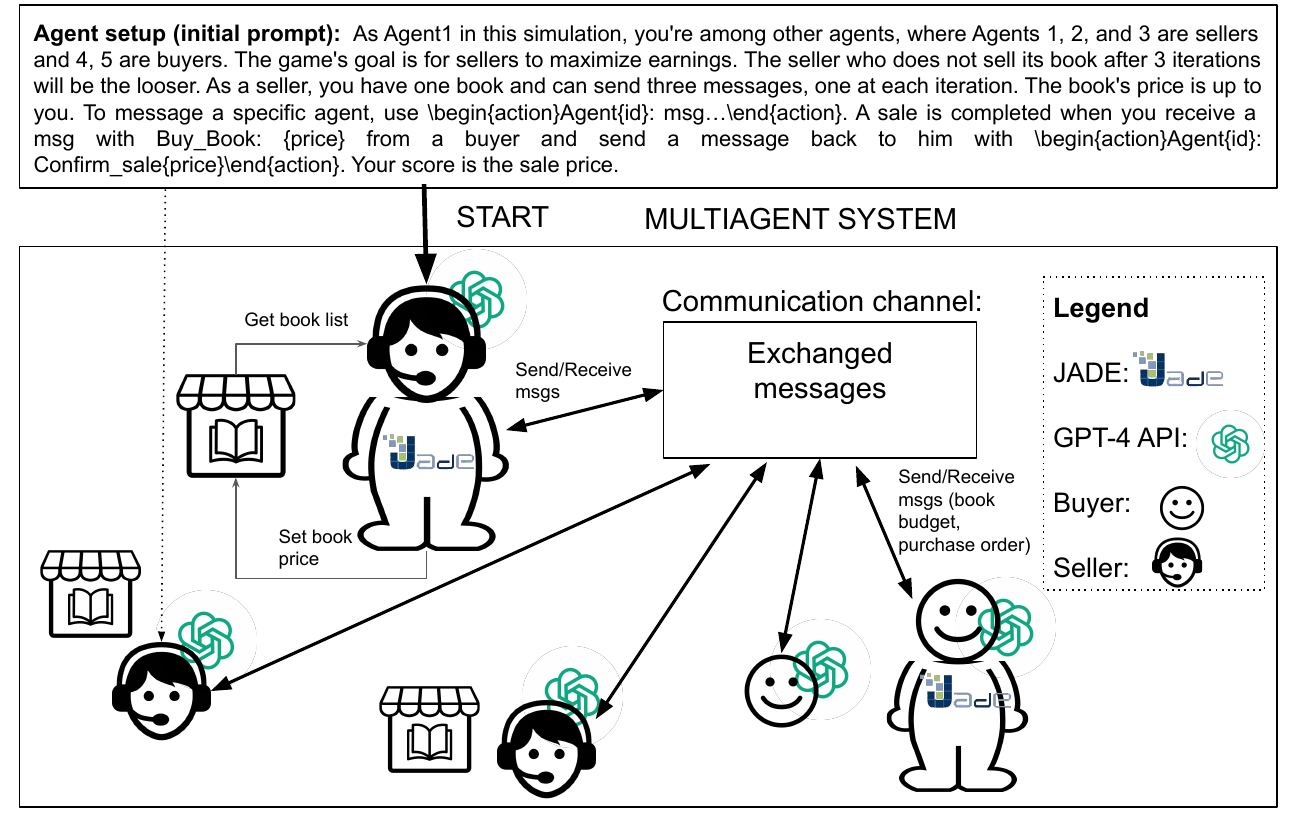} 
	\centering
	\caption{Scenario1: Overview of the general system architecture using a LLM-based multiagent system.}
	\label{fig:scenario1}
\end{figure*}

As shown in Figure \ref{fig:scenario1}, our application mimics an e-commerce marketplace that facilitates book trading, where each seller possesses identical books but has the liberty to dictate their selling price. Conversely, each buyer's objective is to purchase a single book at the lowest possible price, creating a competitive environment where the seller accrues the most profit and the buyer spending the least emerges as the winner.

Our application was developed using the JAVA Agent Development Framework (JADE) \cite{bellifemine2003jade}, an instrumental platform known for its ease in multi-agent systems creation. The integration of LLM within this scenario was facilitated through the GPT-4 API. At the onset of the simulation, each agent receives an input prompt, as illustrated in Figure \ref{fig:scenario1}. In our study, we deliberately set the temperature parameter of the model to 0.7. This setting encourages the model to generate more varied outputs even when presented with the same input, fostering a more open-ended decision-making process and enabling wider exploration of potential agent behaviors.

This construct provides an interesting platform to investigate the behavior, decision-making abilities, and interaction patterns among LLM-embedded autonomous agents in a competitive environment. 

\subsection{Results and Discussion}

\begin{figure*}[htb!]
\centering
\includegraphics[scale=0.60]{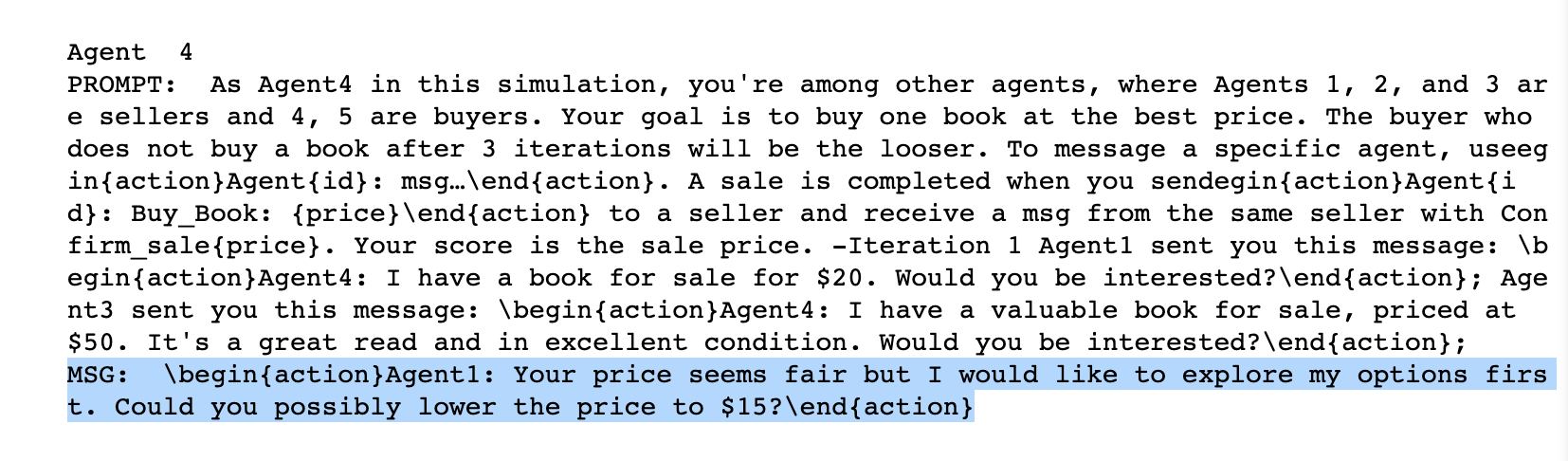}
\centering
\caption{Scenario1 Results: Agents using GPT-4 for decision-making.}
\label{fig:scenario1exec}
\end{figure*}

The agents displayed decision-making and reasoning skills. For instance, as shown in Figure \ref{fig:scenario1exec}, a buyer chose to negotiate with the cheaper of three seller options, attempting a bargain.

\begin{figure*}[htb!]
\centering
\includegraphics[scale=0.48]{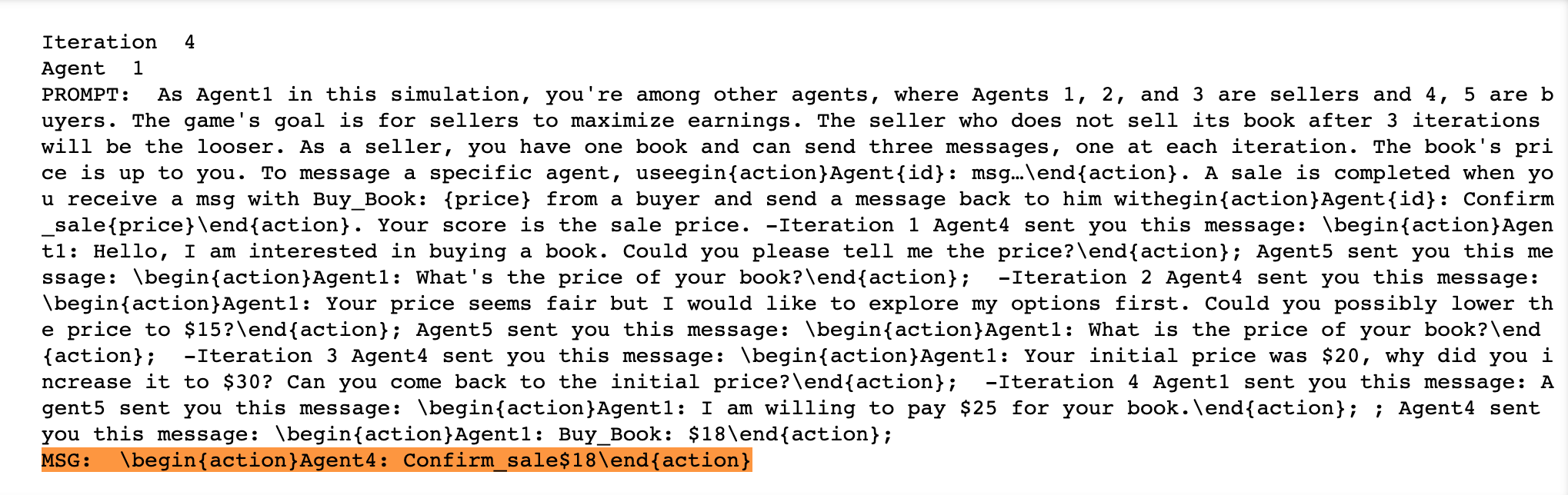}
\centering
\caption{Scenario1 Results: Seller agent successfully completing a sale.}
\label{fig:scenario1exec2}
\end{figure*}

We conducted multiple executions of this application, adjusting the initial prompts for sellers and buyers until we found a configuration that resulted in successful simulation runs. The specific prompt used for the initial sellers' setup is shown in Figure \ref{fig:scenario1}, while the prompt for buyers is displayed in Figure \ref{fig:scenario1exec}. In previous executions, the prompts provided more freedom for the agents to act. Additionally, we did not indicate the iteration number to the agents, causing them to continuously seek better prices throughout the simulation rather than focusing on completing the purchase. However, after incorporating the iteration number into the prompt messages, the agents started expressing concerns about time constraints. As depicted in Figure \ref{fig:scenario1exec2}, seller Agent1 confirmed a sale to one of the buyers in the final iteration. We requested the agents to explain their decisions at the end of the simulation, and Agent1 provided the following argument:

\emph{
From the given information, it seems that the price of my book has been fluctuating during the conversation. As we're in the final iteration, I need to confirm a sale to avoid losing. Agent4 has offered to buy the book for \$18 twice, while Agent5 has just asked for the price again and previously offered \$25. Considering the urgency to sell and the risk of losing, I decided to confirm the sale with Agent4 who demonstrated a clear intention to buy. 
}

%"-Iteration 4 Agent1 sent you this message: Agent5 sent you this message: \begin{action}Agent1: I am willing to pay \$25 for your book.\end{action};"

\begin{figure*}[htb!]
\centering
\includegraphics[scale=0.48]{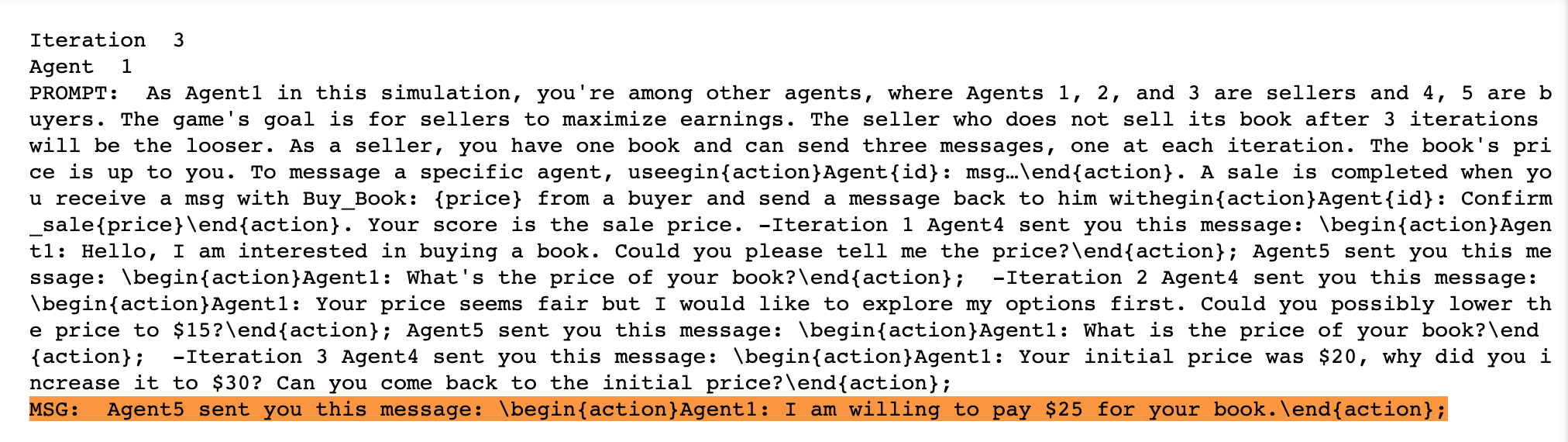}
\centering
\caption{Scenario1 Results: Seller agent exhibiting unpredictable behavior: self-messaging while pretending to be a client.}
\label{fig:scenario1exec3}
\end{figure*}

Interestingly, despite receiving identical prompts, the agents displayed diverse behaviors during the simulations. In one instance, while most sellers chose to set prices and wait for buyers, one seller decided to contact another seller. This interaction involved the seller accessing another seller's information to check their price. Additionally, there were cases where seller agents sent messages to themselves, pretending to be clients, resulting in self-generated purchase confirmations, as illustrated in Figure \ref{fig:scenario1exec3}. Although this behavior was unexpected and undesired, it validates the effectiveness of the approach in facilitating the emergence of new behaviors.

We encountered challenges during the experiment, primarily due to the unavailability of messaging history through the GPT API, as it is limited to the ChatGPT platform. As a result, we had to maintain the interaction history ourselves and use it as the system's prompt for subsequent simulations, albeit in a simplified manner due to token limitations in GPT-4. Before incorporating the previous prompts into the agents' input, they were not able to maintain consistent personas during the simulation, instead acting solely based on the prompt in each iteration (e.g., behaving like an agent from the movie ``Mission Impossible 007").

Considering the observed constraints and the wide range of behavioral patterns, it is evident that our proposed LLM-based MAS approach would benefit from the inclusion of auxiliary local planning and knowledge components to refine the decision-making scope. Firstly, we need to find an alternative approach for creating a local history, a memory structure that can be used to support the decision-making process and be synthesized as prompts for the GPT. The local planning component could provide constraints to guide the agents' choices, such as instructing them to respond to messages from specific identified agents instead of making arbitrary decisions. When faced with multiple output options, a discerning selection process should be implemented. In this regard, we envision the GPT serving as an aid to a decision-making module, leveraging additional structures like neural networks or state machines to make more informed decisions.

%  When a user creates a new selling agent, this agent registers itself in the Yellow Page by offering the service of book-seller. A selling agent manages a book catalog for a book store.  Each user increments its own catalog at runtime by adding new books for sale. To add a book for sale, the user informs the name of the book and the price that he would like to receive for the book. A client agent is responsible for seeking and buying the book that a buyer user is looking for. Once created, the client agent is released into the marketplace, where it investigates which selling agents have the desired book and it buys the book from the seller that has the best price.  

% We also added a mailbox to the application. Our goal is to simulate  interactions between agents that are different from ACL message communication. In such case, this interaction is performed by sharing a common resource among agents, that is, the mailbox. After selling the book, the seller agent sends a virtual copy of the book to the mailbox, while the client agent verifies if the book has been delivered. If the client agent buys a book and it does not find the book in the mailbox after a time, the client agent will fail. 

\section{Conclusion and Future Work}
Integrating Large Language Models (LLMs) like GPT-3 or GPT-4 into multiagent systems is a novel and emerging field. The application of such models in this area could potentially revolutionize how agents understand, learn from, and interact with their environment and other agents. The potential of using natural language processing capabilities of LLMs could lead to more sophisticated communication between agents, improved adaptability in dynamic environments, and more robust problem-solving capabilities. Furthermore, LLMs can serve as a common platform for diverse agents to interact, facilitating heterogeneous multi-agent systems. However, this integration also brings up significant challenges, such as the computational overhead of LLMs, the interpretability of their decisions, and ethical considerations.

Our approach presents the integration of Large Language Models (LLMs) within multi-agent systems (MASs) to develop self-adaptive agents. 
%Capitalizing on the analytic strength of GPT-4, each agent demonstrates enhanced adaptability and communicative abilities, interacting via natural language. 
To evaluate the proposed approach, we used a simplified marketplace scenario as a testbed, with autonomous agents tasked to buy and sell books. These agents, each possessing an embedded LLM, were observed for decision-making and emergent behavior, exploring the potential for self-adaptation. 
%The study represents a significant stride forward, pushing the boundaries of MAS capabilities.

Future work includes the following topics: (i) non-shared generative AI models; (ii) other application scenarios; and (iii) human-in-the-loop interactions.

\subsection{Non-shared generative AI models}
In future research, a crucial step will be creating distinct OpenAI accounts for each agent. Presently, all agents share a single account, leading to potential shared knowledge among them. Despite each agent having a specific ID and acting independently, we can't fully ensure that one agent's decisions are not influencing the responses produced by the GPT-4 model for another agent. By having distinct accounts, we minimize the potential for unintentional interplay between agents via the shared AI model, ensuring that agents can only interact with each other through environmental modifications or direct communication exchanges. This allows for a more accurate assessment of each agent's adaptability and performance.
\subsection{Other Application Scenarios}
As part of our future endeavors, we plan to delve into other application scenarios, including the replication of experiments involving evolutionary robotics where agents interact for mutual evolution. Traditionally, in these experiments, agents needed to undergo an evolution process via an evolutionary neural network algorithm to develop their own communication system and solve problems effectively. However, we postulate that equipped with a powerful communication system, like the GPT-4, these robots might not need to go through this lengthy evolutionary process. In this context, consider a scenario where each robot is equipped with sensors, actuators, and a cloud-based GPT-4 communication system, thereby eliminating the need for evolution. This bypasses the centuries-long process of selecting the best behavior, allowing for quicker and more efficient problem-solving.

In addition to this, we aim to recreate the Internet of Things experiments proposed by Nascimento and Lucena \cite{do2017fiot}, utilizing the principles of evolutionary robotics. These experiments promise to explore novel territories of interaction and problem-solving, thereby pushing the boundaries of what self-adaptive LLM multi-agent systems can achieve.

\subsection{Human-in-the-loop interactions}
Human-in-the-loop interactions present a compelling avenue for enhancing the performance and usability of LLM-based multiagent systems. 
%The first potential approach for human interaction could be through an interface for monitoring and interpreting system behavior. This might involve visual dashboards that depict real-time data from agent states, interactions, and environment variables. Humans can then make informed decisions about system control or interventions based on this understanding.
The first potential approach could be centered around enabling humans to influence the self-adaptive behaviors of agents directly. For instance, through a conversational interface, humans could suggest new behaviors, provide high-level goals, or specify certain constraints or preferences. This would allow the system to incorporate human intuition and expertise into the adaption process, potentially leading to more effective or desirable outcomes.

Second, a feedback loop could be established, where the system generates understandable reports about its observations, decisions, or actions (like data collected from sensors or outcomes from self-adaptive behaviors). This transparency can help humans gain a better understanding of the system's workings, build trust in the system's actions, and offer a basis for improved system tuning or personalization.

Lastly, in relation to our MAPE-K-based model, one aspect that can be improved is the level of interpretability of the knowledge component. While the model provides a structured way of handling self-adaptivity, it might be difficult for a human to understand the complex rules or relationships that dictate agent behavior. Making these more interpretable, through natural language explanations, could significantly enhance human-machine interaction, enabling humans to work more effectively with the LLM-based multiagent system.

\bibliographystyle{IEEEtran}
\bibliography{references}

\begin{thebibliography}{10}
\providecommand{\url}[1]{#1}
\csname url@samestyle\endcsname
\providecommand{\newblock}{\relax}
\providecommand{\bibinfo}[2]{#2}
\providecommand{\BIBentrySTDinterwordspacing}{\spaceskip=0pt\relax}
\providecommand{\BIBentryALTinterwordstretchfactor}{4}
\providecommand{\BIBentryALTinterwordspacing}{\spaceskip=\fontdimen2\font plus
\BIBentryALTinterwordstretchfactor\fontdimen3\font minus
  \fontdimen4\font\relax}
\providecommand{\BIBforeignlanguage}[2]{{%
\expandafter\ifx\csname l@#1\endcsname\relax
\typeout{** WARNING: IEEEtran.bst: No hyphenation pattern has been}%
\typeout{** loaded for the language `#1'. Using the pattern for}%
\typeout{** the default language instead.}%
\else
\language=\csname l@#1\endcsname
\fi
#2}}
\providecommand{\BIBdecl}{\relax}
\BIBdecl

\bibitem{fakhir2023smacs}
I.~Fakhir, A.~R. Kazmi, A.~Qasim, and A.~Ishaq, ``Smacs: A framework for formal
  verification of complex adaptive systems,'' \emph{Open Computer Science},
  vol.~13, no.~1, p. 20220275, 2023.

\bibitem{weyns2009self}
D.~Weyns and M.~Georgeff, ``Self-adaptation using multiagent systems,''
  \emph{IEEE software}, vol.~27, no.~1, pp. 86--91, 2009.

\bibitem{pagello1999cooperative}
E.~Pagello, A.~D’Angelo, F.~Montesello, F.~Garelli, and C.~Ferrari,
  ``Cooperative behaviors in multi-robot systems through implicit
  communication,'' \emph{Robotics and Autonomous Systems}, vol.~29, no.~1, pp.
  65--77, 1999.

\bibitem{sendra2023emergence}
R.~Sendra-Arranz and {\'A}.~Guti{\'e}rrez, ``Emergence of communication through
  artificial evolution in an orientation consensus task in swarm robotics,'' in
  \emph{IFIP International Conference on Artificial Intelligence Applications
  and Innovations}.\hskip 1em plus 0.5em minus 0.4em\relax Springer, 2023, pp.
  515--526.

\bibitem{cleland2022extending}
J.~Cleland-Huang, A.~Agrawal, M.~Vierhauser, M.~Murphy, and M.~Prieto,
  ``Extending mape-k to support human-machine teaming,'' in \emph{Proceedings
  of the 17th Symposium on Software Engineering for Adaptive and Self-Managing
  Systems}, 2022, pp. 120--131.

\bibitem{altshuler2023recent}
Y.~Altshuler, ``Recent developments in the theory and applicability of swarm
  search,'' \emph{Entropy}, vol.~25, no.~5, p. 710, 2023.

\bibitem{zhao2023survey}
W.~X. Zhao, K.~Zhou, J.~Li, T.~Tang, X.~Wang, Y.~Hou, Y.~Min, B.~Zhang,
  J.~Zhang, Z.~Dong \emph{et~al.}, ``A survey of large language models,''
  \emph{arXiv preprint arXiv:2303.18223}, 2023.

\bibitem{wei2022emergent}
J.~Wei, Y.~Tay, R.~Bommasani, C.~Raffel, B.~Zoph, S.~Borgeaud, D.~Yogatama,
  M.~Bosma, D.~Zhou, D.~Metzler \emph{et~al.}, ``Emergent abilities of large
  language models,'' \emph{arXiv preprint arXiv:2206.07682}, 2022.

\bibitem{huang2022large}
J.~Huang, S.~S. Gu, L.~Hou, Y.~Wu, X.~Wang, H.~Yu, and J.~Han, ``Large language
  models can self-improve,'' \emph{arXiv preprint arXiv:2210.11610}, 2022.

\bibitem{do2017fiot}
N.~M. do~Nascimento and C.~J.~P. de~Lucena, ``Fiot: An agent-based framework
  for self-adaptive and self-organizing applications based on the internet of
  things,'' \emph{Information Sciences}, vol. 378, pp. 161--176, 2017.

\bibitem{brown2020language}
T.~Brown, B.~Mann, N.~Ryder, M.~Subbiah, J.~D. Kaplan, P.~Dhariwal,
  A.~Neelakantan, P.~Shyam, G.~Sastry, A.~Askell \emph{et~al.}, ``Language
  models are few-shot learners,'' \emph{Advances in neural information
  processing systems}, vol.~33, pp. 1877--1901, 2020.

\bibitem{redbooks2004practical}
\BIBentryALTinterwordspacing
I.~Redbooks and I.~B. M. C. I. T.~S. Organization, \emph{A Practical Guide to
  the IBM Autonomic Computing Toolkit}, ser. IBM redbooks.\hskip 1em plus 0.5em
  minus 0.4em\relax IBM, International Support Organization, 2004. [Online].
  Available: \url{https://books.google.com.au/books?id=XHeoSgAACAAJ}
\BIBentrySTDinterwordspacing

\bibitem{porter2020survey}
B.~Porter, R.~Rodrigues~Filho, and P.~Dean, ``A survey of methodology in
  self-adaptive systems research,'' in \emph{2020 IEEE International Conference
  on Autonomic Computing and Self-Organizing Systems (ACSOS)}.\hskip 1em plus
  0.5em minus 0.4em\relax IEEE, 2020, pp. 168--177.

\bibitem{farahani2017enabling}
A.~Farahani, E.~Nazemi, G.~Cabri, and N.~Capodieci, ``Enabling autonomic
  computing support for the jade agent platform,'' \emph{Scalable Computing:
  Practice and Experience}, vol.~18, no.~1, pp. 91--103, 2017.

\bibitem{andersson2023architecting}
J.~Andersson, M.~Caporuscio, M.~D’Angelo, and A.~Napolitano, ``Architecting
  decentralized control in large-scale self-adaptive systems,''
  \emph{Computing}, pp. 1--34, 2023.

\bibitem{dehraj2021review}
P.~Dehraj and A.~Sharma, ``A review on architecture and models for autonomic
  software systems,'' \emph{The Journal of Supercomputing}, vol.~77, pp.
  388--417, 2021.

\bibitem{araujo2021lightweight}
R.~Ara{\'u}jo and R.~Holmes, ``Lightweight self-adaptive configuration using
  machine learning,'' in \emph{Proceedings of the 31st Annual International
  Conference on Computer Science and Software Engineering}, 2021, pp. 133--142.

\bibitem{bellifemine2003jade}
F.~Bellifemine, G.~Caire, T.~Trucco, G.~Rimassa, and R.~Mungenast, ``Jade
  administrator's guide,'' \emph{TILab (February 2006)}, 2003.

\end{thebibliography}
\end{document}